\documentclass[12pt,preprint]{aastex}

\newcommand\plotoner[1]{%
 \centering 
 \leavevmode 
 \includegraphics[angle=270,width={\columnwidth}]{#1}%
}%

\begin{document}

\shorttitle{High time resolution of the Vela glitch}
\shortauthors{Dodson et al.}

\title{High time resolution observations of the January 2000 glitch in
the Vela pulsar}

\author{R.G. Dodson\altaffilmark{1} and
  P.M. McCulloch\altaffilmark{1} and
  D.R. Lewis\altaffilmark{1}}

\affil{School of Mathematics and Physics, University of Tasmania}

\altaffiltext{1}{School of Mathematics and Physics, University of Tasmania, 
     GPO Box 252-21, Hobart, Tasmania 7001, Australia}

\email{Richard.Dodson@utas.edu.au}

\begin{abstract}
  Pulsars are rotating neutron stars, sweeping the emission regions
  from the magnetic poles across our line of sight. Isolated neutron
  stars lose angular momentum through dipole radiation and (possibly)
  particle winds, hence they slow down extremely steadily, making them
  amongst the most reliable timing sources available. However, it is
  well known that younger pulsars can suffer glitches, when they
  suddenly deviate from their stable rotation period. On 2000 January
  16 (MJD 51559) the rate of pulsation from the Vela pulsar (B0833-45)
  showed such a fractional period change of {\rm $3.1\times 10^{-6}$},
  the largest recorded for this pulsar. The glitch was detected and
  reported by the Hobart radio telescope. The speedy announcement
  allowed the X-ray telescope, Chandra, and others, to make Target of
  Opportunity observations. The data placed an upper limit of 40
  seconds for the transition time from the original to the new
  period. Four relaxation timescales are found, which are believed to
  be due to the variable coupling between the crust and the interior
  fluid.  One is very short, about 60 seconds; the others have been
  previously reported and are 0.56, 3.33 and 19.1 days in length.
\end{abstract}



\section{Introduction}

Observations of pulsar glitches, in addition to providing insights
into the phenomenon itself, offer one of the few probes of neutron
star structure, and thus the physics of ultra-dense matter. Vela is
the brightest known radio pulsar, and as it is at a declination of
-45$^o$, it is above the horizon at the Hobart Radio Observatory
(Mount Pleasant) for more than 18 hours a day.  It undergoes large
glitches in pulse rate every few years and so provides an excellent
probe of the physics of glitches.

Since 1981 the University of Tasmania has devoted a 14m diameter
antenna at its Mt Pleasant Observatory to measurements of arrival
times of pulses from the Vela pulsar. During this time we have
observed 7 large glitches, or sudden decreases, in the period of the
pulsar \citep{vela_1983,vela_1987,christ_nat,ppp_vela}.

The telescope has a single pulse observing system whose speed and
sensitivity have been enhanced in order to answer a number of
questions; how quickly does the crust accelerate to the new period
during a glitch, how soon does the recovery from the glitch start,
and what is the form of this recovery?

On 2000 January 16 (MJD 51559) the rate of pulsation jumped with a
fractional period change of {\rm $3.1\times 10^{-6}$}, the largest
recorded for this pulsar. The glitch was automatically detected and we
issued an IAU telegram \citep{iau_7347} within 12 hours, allowing the
X-ray telescope, Chandra, to make Target of Opportunity (TOO) observations
\citep{vela_xray_pwn}. These observations have so far failed to find
the signature of neutron star heating, which was the driver for the
TOO, but have produced spectacular images of the X-ray pulsar
wind nebula.

\section{Observations}

The reported observations were made simultaneously at 635~MHz, 990~MHz
and 1390~MHz, to allow continuous measurement of the dispersion
measure (DM). Dual channel receivers with an equivalent system
temperature of 60 K are used at each frequency, except for 1390~MHz,
where only the right handed circular polarisation is observed. Receiver
bandwidths of 250 kHz at 635~MHz, 800~kHz at 990~MHz and 2~MHz at
1390~MHz limit the pulse broadening from interstellar dispersion to
less than 1\% of the pulse period. The detected output from each
receiver was folded for 2 minutes to give integrated pulse profiles
from 1,344 pulses. The profiles were subsequently combined to give
profiles of total intensity. The signal-to-noise ratio in each total
intensity profile is typically 30:1, allowing a mean pulse arrival
time to be determined to an accuracy of $80\mu$s at 635~MHz, $60\mu$s
at 990~MHz and $180\mu$s at 1390~MHz per integration.

The recent improvements in the time resolution have been achieved by
incoherently de-dispersing over 8 adjacent channels at 990~MHz,
thereby increasing the signal to noise ratio and allowing observations
of single pulses. The de-dispersed bandpass is sampled at 2 kHz and
recorded directly onto disk for later retrieval, while an `on the fly'
monitoring system folds on a ten second basis for which the RMS is
$85\mu$s. These arrival times are monitored and if a glitch is
detected a warning is issued and the single pulse data are retained.

\section{Timing fits}
The full, folded, timing profiles are stored in EPN format \citep{epn},
cross correlated with high signal to noise profile templates, then fitted 
to find the time of arrival (TOA) of the pulse. 

A Taylor expansion is parameterised in terms of the spin-down model
phase ($\phi$) around a selected epoch ($t$ relative to $t_0$) with
frequency $\nu$, and frequency derivative $\dot{\nu}$, and has the
form;

\begin{equation}
\phi_m(t)=\phi_0+\nu (t-t_0)+\frac{1}{2}\dot{\nu} (t-t_0)^2+
\frac{1}{6}\ddot{\nu} (t-t_0)^3
\end{equation}

The post glitch frequency change can be written as the sum of the
pre-glitch frequency plus the permanent change at the time of the
glitch, plus other decay terms:

\begin{equation}
\nu(t) = \nu_0 + \Delta \nu_p+ \Delta \dot{\nu_p} t+ \frac{1}{2}\Delta
\ddot{\nu_p} t^2+
\Sigma_n \Delta \nu_n e^{-t/\tau_n}
\end{equation}

However, it should be noted that some investigators set the permanent
change terms, $\Delta \dot{\nu_p}$ and $\Delta \ddot{\nu_p}$ to zero.
We have allowed $\Delta \dot{\nu_p}$ to take non-zero values, but keep
$\Delta \ddot{\nu_p}$ as zero.  Both of these terms affect only the
longest time-scales, where both timing noise and the occurrence of the
subsequent glitches make definitive separation impossible.

The data presented here were recorded between MJD 51505 and 51650.
The arrival time data have been transformed to the Solar System
barycentre using standard techniques.  The position and proper motion
of the Vela pulsar was defined by data from the Radio VLBI position of
\citet{legge_private}. The recorded TOAs from all frequencies and both
systems were fitted in the program {\bf TEMPO} \footnote{See
http://pulsar.princeton.edu/tempo (Taylor, Manchester, Nice, Weisberg,
Irwin, Wex, \& Standish 1970)}. The results of this fit are given in
Table 1.

Figure 1a shows the residuals from the pre-jump fit for data taken on
2000 January 16 (MJD 51559) with the full polarisation system.
Shortly after 07:34 UT, the residuals diverge from the fit, indicating
a sudden decrease in pulse period.  Figure 1b shows an hour of data
starting at MJD 51559.3. Individual data points represent ten second
averages constructed from the single pulse data. The period jump occurs
on a very short timescale, without warning. The observations are
consistent with an instantaneous change in period; modeling has shown
that a spin-up timescale of forty seconds would produce a three sigma
signal.

The separation into four time-scales is clear. The longer three decay
terms are similar to those previously reported
\citep{alpar_1993,flanagan}, and are in an approximately equal ratio
of 5.9:5.7. These have been associated with the vortex creep models by
\citet{alpar_1993} and others. The fast decay timescale, not
previously observed (or observable) is shown separated from the other
effects in figure 2. We have subtracted the terms found by {\bf TEMPO}
in the 2 minute data from the single pulse data folded for 10
seconds. In this plot a gradual spin-up (as opposed to instantaneous)
would be a negative excursion around the projected time of the glitch,
as we'd have overestimated the phase in the model.
We see a positive excursion, followed by a rapid decay. A positive
excursion could be produced by the pulsar slowing down just before the
glitch or, more likely, if the estimate of the glitch epoch was too early
because there was an extra component not resolvable in the 2 minute
data. Once the original fit was removed this would give a linear rise
with the gradient $\Delta \nu$ found in the 2 minutes data,
followed by a decay. We have fitted this rise ($\Delta \nu \Delta t$)
followed by a forth exponential decay term to give the true glitch
epoch and the fastest decay term.

\section{The single pulse system}

Since the acceleration of the crust cannot be instantaneous, it should
be possible to observe the spin-up of the rotation period of the
pulsar. The parallel single pulse system designed to observe this has
a three $\sigma$ detection limit of less than 40 seconds. This is in
contrast to the observations made on the Crab \citep{crab_spinup},
where the spin-up timescale has been observed to take about half a
day. No variation could be seen either before or after the glitch in
profile shape, magnitude, or polarisation in the radio observations on
any time-scale. The raw time series using both long and short
timescales has been closely examined to look for other periodic and
quasi-periodic signals, but no extra terms were found.

\section{Estimates for the force from the vortex lines}

We will use our data to explore one model of pulsar glitches in which
the glitches are a consequence of pinning of superfluid vortices to
the stellar crust. As the neutron star slows due the external drag on
the crust, an angular momentum excess develops in the pinned
superfluid. Ultimately the vortices unpin catastrophically,
transferring the excess momentum to the crust, producing a glitch.

Our high time resolution studies permit constraints on the manner in
which the liquid interior imparts its angular momentum to the
crust. This coupling is thought to be due to dissipative motion of
vortices through the crustal lattice after unpinning.  Various papers
have tackled calculation of the drag on vortex lines within the
superfluid (e.g. \citet{bildsten_drag,epstein_baym_92,jones_92})
produced by vortex-neutron scattering in the crust. We will ignore the
details of the interaction here, as it is not a simple
process. However, given that we have a lower limit for the spin-up
time, we can obtain an observational limit for the force per length,
$f$, that a vortex line exerts on the superfluid during this spin-up
phase in Vela. Assuming that pinned vorticity in the crust is
responsible for the glitch, it is straightforward to calculate the
torque required to spin-up the pulsar, and then, by assuming canonical
neutron star radii, crust depths and vortice lengths, give the force
per unit length.

During the spin-up, the torque on the crust, ${\cal N}$, is:

\[ {\cal N} = I_s \Delta\Omega_s/\Delta t \]

where $I_s$ is the moment of inertia of the portion of the superfluid that
drives the glitch and $\Delta\Omega_s$ is the change in angular
velocity of that component, $\Delta t$ is the spin-up time. Since
angular momentum is conserved in the glitch, the torque is

\[ {\cal N} = I_{crust} \Delta\Omega/\Delta t \]

where $I_{crust}$ is the moment of inertia of the solid crust plus any
component that is coupled to it over a timescale in less than the
spin-up time $\Delta t$, and $\Delta\Omega$ is the observed spin-up of
the crust over $\Delta t$.

In magnitude, the torque is

\[ {\cal N} = F_\phi R \]

where $F_\phi$ is the azimuthal component of the force that vortices
exert on the lattice during the glitch, and $R$ is the average
distance of vortices from the rotation axis, approximately the stellar
radius. The azimuthal force is

\[ F_\phi = f H N_v \]

where $f$ is the force exerted per unit length per vortex, $H$ is the
average length of a vortex, and $N_v$ is the total number of vortices
that are participating in the spin-up. 

The number of vortices is

\[ N_v = n_v  2\pi R \Delta R \]

where $n_v$ is the number of vortices per unit area and $\Delta R$ is
the thickness of the pinning region (approximately the thickness of
the crust). The number density of vortices in a rotating superfluid is

\[ n_v = 2\Omega_s/\kappa \]

where $\Omega_s$ is the superfluid rotational velocity, approximately
equal to the rotational velocity of the crust and $\kappa$ is the
quantum of circulation of a neutron fluid (0.0019 cm$^2$/s). Putting
these expressions together gives a constraint for $f$:

\begin{equation}
f = I_{crust} \frac{\kappa \Delta\Omega/\Delta t}
                       {H 4\pi \Omega R^2 \Delta R}
\end{equation}

$\Delta R$, the thickness of the crust, is about 0.1 km. We take $H$,
the length of a vortex, to be roughly $\Delta R$.

We defined $I_{crust}$ as the moment of inertia of the solid plus any
other component coupled to it in a timescale less than $\Delta t$. If
the core fluid is tightly coupled to the solid crust over $\Delta t$,
then $I_{crust}$ is approximately the moment of inertia of the entire
star (about $10^{45}$ g~cm$^2$). On the other hand, if the core is
decoupled over $\Delta t$ then $I_{crust}$ would correspond to the
moment of inertia of the solid crust only, which is about $10^{43}$
g~cm$^2$ for a moderate equation of state. So we can obtain a lower
limit on $f$.

\[ f \geq 10^{12} \ I_{crust,43}\ {\rm dyn/cm} \]

Where $I_{crust,43}$ is the crust inertia in units of $10^{43}$
g~cm$^2$. Vortices could pin to the lattice with a force $\sim 10^4$
times this \citep{lc_pinning_2001}, but our observations constrain the
value of $f$ during the dissipative regime, whilst the pulsar is being
spun up. Putting the same parameters into the Crab time-resolved
glitch of MJD 50259.93 \citep{crab_spinup} we find that in that case
$f \approx 4 \times 10^{5}$~dyn/cm. This is further evidence, if more
was needed, that the Crab pulsar is a very different object from the
Vela pulsar. The major difference between the Crab and the Vela pulsars is
the age, and therefore the temperature. \citet{jones_92} found a
strong temperature dependence for some scattering processes, and these
differences may be key to our understanding of the momentum transfer
in glitches. Further spin-up observations for both the Crab and Vela
are required.


\section{Future development}

Further improvements of the single pulse system are being undertaken,
with a new coherent dedispersion system, which will allow a detection
limit of a few seconds, the order of the fastest coupling times in all
theoretical models. This should also allow us to measure $f$, the
force per unit length on the vortices, and confirm the fast decay
term. Higher time resolution will allow further constraints on the
coupling of the crust to the liquid interior, including the core.
We will also co-ordinate with Chandra to reduce the TOO delay time in
order to determine whether a very short lived thermal pulse might be
emitted.

\vspace{2cm}
\noindent

\acknowledgments{The day to day maintenance of the 14m telescope, over
20 years has been a herculean task, to which too many people to name
have contributed. Also we greatly appreciated the extremely helpful
suggestions on the text from the referee, B. Link and also from
S. Johnston.}

\newpage
\begin{table}[h]
\sffamily
\begin{center}
\begin{tabular}{|c|c|c|}
\hline
\multicolumn{3}{|c|}{Parameters for Epoch 51559}\\
\hline
$\nu/Hz$ & $\dot{\nu}/Hz~s^{-1}$ & $\ddot{\nu}/Hz~s^{-2}$ \\
\hline
11.194615396005 & -1.55615E-11 & 1.028E-21 \\
\hline
\end{tabular}

\begin{tabular}{|l|l|}
$\Delta \nu_p/Hz$ & $\Delta \dot\nu_p/Hz~s^{-1}$ \\
\hline
3.45435(5)E-05&-1.0482(2)E-13\\
\hline
$\tau_n$ & $\Delta \nu_n/10^{-6} Hz$\\
\hline
$1.2\pm0.2$mins & 0.020(5)\\ 
00.53(3) days& 0.31(2)\\ 
03.29(3) days& 0.193(2)\\ 
19.07(2) days& 0.2362(2)\\ 
\hline
DM & 67.99 \\
\hline
\end{tabular}
\end{center}
\caption{Parameters for the glitch epoch 51559.3190. Errors are the
one sigma values.  The data fit is from MJD 51505 through to 51650
(November 1999 to April 2000).}

\label{tab:parameters}
\end{table}

\plotone{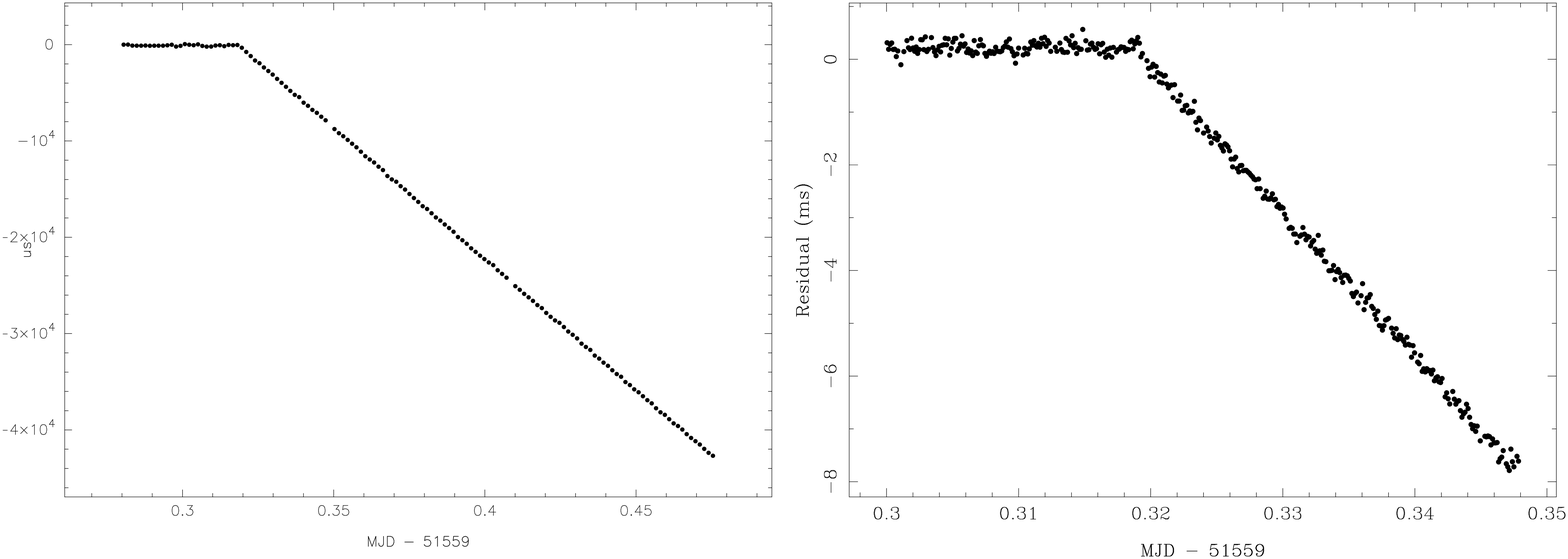}
\figcaption{Arrival time residuals from the pre-glitch model for both
the three frequency (1a) and the single pulse (1b) systems. Figure 1a
has folds of two minutes and figure 1b has folds of ten seconds.}
\newpage
\plotoner{fast_decay_alt.eps}
\figcaption{The previously unobserved fast decay, 
with ten second folds. The other decay terms are removed revealing the later 
start, and decay of the fastest term.} 
\newpage

\end{document}